\documentstyle[11pt]{article}
\begin{document}
\newcommand{\be}{\begin{equation}}
\newcommand{\ee}{\end{equation}}
\def\theequation{\arabic{section}.\arabic{equation}}
\begin{titlepage}
\title{Energy conditions and classical scalar fields}
\author{S. Bellucci$^1$ and V. Faraoni$^{1,2}$\\ \\
{\small \it $^1$ INFN-Laboratori Nazionali di Frascati, 
P.O. Box 13, I-00044 Frascati, Roma, Italy}\\
{\small \it $^2$ Physics Department, University of Northern British 
Columbia}\\
{\small \it 3333 University Way, Prince George, B.C., V2N 4Z9, Canada}\\
}
\date{}
\maketitle
\thispagestyle{empty}      \vspace*{1truecm}
\begin{abstract}
Attention has been recently called upon the fact that the weak and null 
energy conditions and the second law of thermodynamics are   
violated in wormhole solutions of Einstein's theory with classical, 
nonminimally coupled, scalar fields as material source. It is shown that the 
discussion is only meaningful when ambiguities in the definitions of 
stress-energy tensor and energy density of a nonminimally coupled scalar 
are resolved. The three possible approaches are discussed with emphasis 
on the positivity of the respective energy densities and covariant 
conservation laws. The root of the ambiguities is traced to the energy 
localization problem for the gravitational field.
\end{abstract} \vspace*{0.3truecm} 

\end{titlepage}

\section{Introduction}
\setcounter{equation}{0}

Scalar fields are ubiquitous in particle physics and in cosmology:
examples are the Higgs boson of the Standard Model, the string dilaton,
the superpartner of spin $1/2$ particles in supergravity, and 
the Brans-Dicke
field \cite{BD}; or the inflaton field and the quintessence scalar field in
cosmology \cite{KolbTurner,NMCquintessence}. 

It is natural that a scalar field $\phi$ living in a curved space couple
explicitly to the Ricci curvature $R$ of spacetime, as described by the
action\footnote{We adopt the notations and conventions of
Ref.~\cite{Wald}; in particular, conformal coupling corresponds to 
$\xi=+1/6$ in these notations.} 
\be 
S = \int d^4 x \sqrt{-g} \left[ \left(
\frac{1}{2\kappa}-\frac{\xi \phi^2}{2} \right) R-\frac{1}{2} g^{ab}
\nabla_a \phi \nabla_b \phi -V( \phi ) \right] + S_{m} \; , \label{1} 
\ee
where $\kappa \equiv 8\pi G $ ($G$ being Newton's constant), $V( \phi)$ is
the scalar field potential, the dimensionless coupling constant $\xi$
describes the explicit nonminimal coupling (hereafter referred to as NMC),
and $S_m$ is the action for all forms of matter different from $\phi$. 

NMC is introduced by first loop corrections: even if a classical scalar
field theory is minimally coupled (i.e. $\xi=0$), first loop corrections
shift the value of the coupling $\xi$ to a nonzero one
\cite{FordToms82,rengroup}, typically of order $10^{-1}$-$10^{-2}$
\cite{AllenIshihara}. A nonzero $\xi$ is also required by independent
arguments in specific high energy theories (see
\cite{Faraonihep,FaraoniPRD96} for reviews). Even at the classical level,
and in a minimalist's point of view restricted to general relativity, the
Einstein equivalence principle \cite{Will} forces the coupling to be
nonminimal.\footnote{The principle was advocated in a
proposal of aconsistent quantum gravity \cite{bs}.
For the consequences of the tests of the
principle in a specific framework, see e.g.
\cite{new1}.} More precisely, the consideration of a conformal
coupling (i.e. $\xi=1/6$) is required in this case
\cite{SonegoFaraoni,GribPoberiiRodrigues}.
\footnote{In a different context, superconformal
invariance was proposed to explain why newtonian gravity shadows
cosmological constant effects \cite{new3}.}

In short, NMC is an unavoidable feature of scalar field physics in curved
spaces. Even without knowledge of this fact, it would be wise to
incorporate NMC in a scalar field theory due to the fact that NMC can have
a dramatic effect on the solutions of the theory and the physics 
that they describe. As an example, consider the well known chaotic 
inflationary
scenario with the potential $V=\lambda \phi^4$; the inclusion of NMC with
a value of $\xi$ as small as $10^{-3}$ prevents the existence of
inflationary solutions unless an unacceptable fine-tuning
 is allowed, thus ruining an inflationary scenario which is very successful
for $\xi=0$ \cite{FutamaseMaeda}.  NMC is relevant in cosmology (during
inflation and the quintessence era) \cite{NMCcosmology}, quantum
cosmology \cite{quantumcosmology}, and in the physics of boson stars 
\cite{Jetzer} and
wormholes \cite{Visserbook}.  Recently\footnote{See Ref.~\cite{Hiscock}
for an earlier report of this phenomenon.}, it was pointed out that the
innocent-looking theory described by the action (\ref{1}) leads to
negative energy fluxes \cite{BarceloVisserCQG}, opening the door to exotica
such as traversable wormholes \cite{BarceloVisserCQG}, warp drives
\cite{warpdrives}, and time machines \cite{Visserbook}. Even worse, the
second law of thermodynamics is put in jeopardy by the theory (\ref{1})
\cite{FordRomanpreprint}. Note that the system described by eq.~(\ref{1})
does not have quantum nature, such as those often considered in the
literature and for which violations of the weak and null energy conditions
are temporary but do not persist on average; the theory (\ref{1}) is,
instead, a classical one.

The negative energy problem is actually much more general than the theory
(\ref{1}): it appears in Brans-Dicke and scalar-tensor theories, in higher
derivatives theories of gravity, in the low energy limit of string
theories and also in brane world theory\footnote{This is not surprising,
since the Randall-Sundrum model somehow reduces to a Brans-Dicke theory
\cite{GarrigaTanaka,BarceloVisserhep}, which suffers from negative 
kinetic energies.}. In this paper, however, we mainly deal with the NMC 
case. 

It turns out that, with NMC, there are ambiguities in the definition of
the scalar field stress-energy tensor and, as a consequence, in the
definition of energy density, pressure, and equation of state. In order to
be able to speak of energy conditions, a specific choice of definitions of 
energy density and pressure must be made; this need is made clear in 
Sec.~2 of this 
paper. In addition, conservation laws for all the possible versions of
stress-energy tensors (there are three of them) are discussed. 

In Sec.~3, the violation of weak and null energy
conditions is discussed, and it is pointed out that these violations are 
not universal, 
i.e. there are physically  interesting situations in which the energy density
(defined in a meaningful way) cannot be negative even in the presence of
NMC. There are also situations in which negative energies are possible but
do not lead to runaway solutions and dynamical catastrophes. And there are
situations, which have been
 pointed out in the literature, in which negative energies become
embarassing for a theory with a generalized coupling
 between the scalar and the gravitational fields \cite{FGN}.\footnote{
For a low-energy theorem in gravity coupled to scalar matter, see. e.g.
\cite{abs}.} A possible  cure, albeit
a drastic one, for the problem of negative energies with NMC is pointed
out in Ref.~\cite{FordRomanpreprint} (see also the references in
\cite{FGN}), and consists in the reformulation of the theory in the
Einstein conformal frame, on the lines of what is done for scalar-tensor
theories \cite{FGN}. This possibility is not surprising since the action
(\ref{1}) can be rewritten as a scalar-tensor theory by means of a
suitable scalar field redefinition. Whether this way out of the problem is
needed, or at all satisfactory, will be left to the judgment of the reader. 


\section{Ambiguities in the stress-energy tensor}
\setcounter{equation}{0}

In this section we discuss the three possible and inequivalent ways of
writing the field equations, the corresponding ambiguity in the definition
of the energy-momentum tensor of the scalar field (and therefore of the
energy density, pressure, and equation of state), and the corresponding
covariant conservation properties in the presence of NMC. 

The field equations derived by varying the action (\ref{1}) are \be
\label{2} G_{ab} = \kappa \left[ \nabla_a \phi \nabla_b \phi -\frac{1}{2}
g_{ab} \nabla^c \phi \nabla_c \phi -V( \phi) \, g_{ab} +\xi \left( g_{ab}
\Box -\nabla_{a}\nabla_{b} \right) ( \phi^2 ) +\xi \phi^2 G_{ab}
+T_{ab}^{(m)} \right] \; , \ee 
where 
\be \label{3} T_{ab}^{(m)} =-\,
\frac{2}{\sqrt{-g}}\, \frac{\delta S_m}{\delta g^{ab}} 
\ee 
is the usual
stress-energy tensor of ordinary matter (different from $\phi$), which
does not explicitly depend on $\phi$.  There are now three ways to
proceed, corresponding to different ways of rewriting the field equations
(\ref{2}). 

\subsection{Procedure \`a la Callan-Coleman-Jackiw}

This procedure follows the one originally adopted by Callan, Coleman and
Jackiw \cite{CCJ} for conformal coupling. NMC is best known after this
work, although it had been introduced earlier in a different context
\cite{ChernikovTagirov}.  One writes the Einstein equations (\ref{2}) as
\be \label{4} 
G_{ab}=\kappa \left( T_{ab}^{(I)}+T_{ab}^{(m)} \right) \; ,
\ee 
where 
\be \label{5} 
T_{ab}^{(I)}= \nabla_a \phi \nabla_b \phi
-\frac{1}{2} g_{ab} \nabla^c \phi \nabla_c \phi -V( \phi) \, g_{ab} +\xi
\left( g_{ab} \Box -\nabla_{a}\nabla_{b} \right) ( \phi^2 ) +\xi \phi^2
G_{ab} 
\ee 
contains the ``geometric'' contribution $\xi\phi^2 G_{ab}$ incorporating 
the Einstein tensor $G_{ab}$.

Let us discuss the conservation properties of $T_{ab}^{(I)}$: the contracted Bianchi identities 
$\nabla^b G_{ab}=0$ yield
\be \label{6}
\nabla^b T_{ab}^{(I)} +\nabla^b T_{ab}^{(m)}=0 \; .
\ee
Since $ T_{ab}^{(m)}$ does not depend on $\phi$ and $\nabla^b T_{ab}^{(m)}$  
vanishes when the field $\phi$ is switched off, one concludes that $T_{ab}^{(m)}$ and $T_{ab}^{(I)}$
are covariantly conserved separately,
\be \label{7}
\nabla^b T_{ab}^{(m)}= 0 \; ,
\ee
\be \label{8}
\nabla^b T_{ab}^{(I)}= 0 \; .
\ee
The energy density 
measured by an observer with four-velocity $u^c$ ($u_cu^c=-1$) is, in this
approach,
\be \label{8bis}
\rho_{(total)}^{(I)} \equiv \left[ T_{ab}^{(I)} +T_{ab}^{(m)} \right]u^au^b=
\rho^{(I)}_{\phi}+\rho^{(m)}\;,
\ee
where
\be \label{8ter}
\rho_{\phi}^{(I)}=T_{ab}^{(I)} u^a u^b=
\left( u^c \nabla_c \phi \right)^2  +\frac{1}{2} 
\nabla^c \phi \nabla_c \phi + V( \phi) 
-\xi \Box ( \phi ^2) -\xi u^a u^b \nabla_{a}\nabla_{b} ( \phi^2 ) 
+\xi \phi^2 G_{ab}u^au^b \; ,
\ee
and $\rho^{(m)}=T_{ab}^{(m)}u^a u^b$ as usual. The trace of $T_{ab}^{(I)}$ reduces to 
\be
T^{(I)}=-\partial^c \phi \, \partial_c \phi -4V( \phi ) +3\xi \Box (\phi^2) 
-\xi R \phi^2 \;. 
\ee

\subsection{The effective coupling approach}

The second way of proceeding is to rewrite the field equations (\ref{2})
by taking the term $ \kappa \xi \phi^2 G_{ab} $ to their left hand side,
\be \label{9} 
\left( 1-\kappa \xi \phi^2 \right) G_{ab}=\kappa \left(
T_{ab}^{(II)} + T_{ab}^{(m)} \right) \; , 
\ee 
where 
\be \label{10}
T_{ab}^{(II)}= \nabla_a \phi \nabla_b \phi -\frac{1}{2} g_{ab} \nabla^c
\phi \nabla_c \phi -V( \phi) \, g_{ab} +\xi \left( g_{ab} \Box
-\nabla_{a}\nabla_{b} \right) ( \phi^2 ) = T_{ab}^{(I)} -\xi \phi^2 G_{ab}
\; , 
\ee 
and to further divide eq. (\ref{9}) by the factor
$1-\kappa\xi\phi^2$, obtaining 
\be \label{11} 
G_{ab}=\kappa_{eff} \left(
T_{ab}^{(II)}+T_{ab}^{(m)} \right) \; , \ee where \be \label{Geff}
\kappa_{eff}( \phi) \equiv \frac{\kappa}{1-\kappa \xi \phi^2} \ee is an
effective gravitational coupling for {\em both} $T_{ab}^{II)} $ and $
T_{ab}^{(m)}$. This procedure is analogous to the familiar identification
of the Brans-Dicke scalar field $\phi_{BD}$ with the inverse of an
effective gravitational constant ($G_{\phi}=\phi_{BD}^{-1}$) in the
gravitational sector of the Brans-Dicke action 
\be 
S_{BD}=\int d^4 x \sqrt{-g} \left( \phi_{BD} R -\frac{\omega}{\phi_{BD}} 
\nabla^a \phi_{BD} \, \nabla_a \phi_{BD} \right) \; . 
\ee 
It is clear that the division by the factor
$1-\kappa\xi\phi^2$ leads to loss of generality for $\xi>0$; solutions of
eq. (\ref{2}) with scalar field attaining the critical values 
\be \label{13} 
\pm \phi_c \equiv \pm \frac{1}{\sqrt{\kappa\xi }}
\;\;\;\;\;\;\;\;\;\;\;\;\;\;\;\;\; \left( \xi > 0 \right) \; , 
\ee 
are  missed when considering
eq.~(\ref{11}) in the effective coupling approach. For example, solutions
in which $\phi$ is constant and equal to the critical values (\ref{13})
appear in models of wormholes \cite{BarceloVisserCQG} and as special
heteroclinics in the phase space of scalar field cosmology
\cite{Gunzigetalpreprint} (see the Appendix for a discussion of solutions
with constant scalar field). In the effective coupling approach these
solutions, which are important ones for wormholes and cosmology, are missed 
and can only be recovered
by going back to the primitive form (\ref{2}) of the field equations. In
other words, eq.~(\ref{11}) is less general than eq.~(\ref{2}). 

Let us discuss now the conservation laws for $T_{ab}^{(II)}$; the
contracted Bianchi identities imply that \be \label{14} \nabla^b \left(
T_{ab}^{(II)}+T_{ab}^{(m)} \right)= \frac{-2\kappa \xi \phi}{1-\kappa \xi
\phi^2} \,\left( \nabla^b \phi \right) \left( T_{ab}^{(II)} +T_{ab}^{(m)}
\right) \; .  \ee By reasoning as in the previous case, one obtains
$\nabla^b T_{ab}^{(m)}=0$ and \be \label{15} \nabla^b T_{ab}^{(II)}=
\frac{-2\kappa \xi \phi}{1-\kappa \xi \phi^2} \, \left( \nabla^b \phi
\right) \left( T_{ab}^{(II)}+T_{ab}^{(m)} \right) \; . \ee One can have a
better idea of the consequences of the conservation law (\ref{15}) by
considering ordinary matter in the form of a dust fluid with corresponding
energy-momentum tensor $T_{ab}^{(m)} =\rho v_a v_b $, where $v^c$ is the
dust four-velocity. Eq.~(\ref{15}) then yields 
\be \label{16} 
\left( \frac{d\rho^{(m)}}{d\lambda} +\rho^{(m)} \nabla^b v_b 
+\frac{2\kappa \xi \rho^{(m)}\phi}{1-\kappa \xi \phi^2} \, 
\frac{d\phi}{d\lambda} \right)
v_a + \rho^{(m)} \, \frac{D v_a}{D\lambda}=0 \; , 
\ee 
where $\lambda $ is
an affine parameter along the worldlines of fluid particles.
Eq.~(\ref{16}) is decomposed into the geodesic equation 
\be \label{17}
\frac{Dv^a}{ D\lambda} \equiv v^b \nabla_b v^a = \frac{d^2 x^a}{d\lambda^2}+
\Gamma^a_{bc}\frac{dx^b}{d\lambda}\frac{dx^c}{d\lambda}=0 
\ee 
and the modified conservation equation 
\be \label{18}
\frac{d\rho^{(m)}}{d\lambda}+\rho^{(m)} \nabla^b v_b +\frac{2\kappa \xi
\phi \, \rho^{(m)}}{1-\kappa \xi \phi^2} \,\, \frac{d\phi}{d\lambda} = 0
\; . \ee 
Test particles move on geodesics, thus verifying the geodesic
hypothesis\footnote{This is not the case, considered later in this paper,
of Brans-Dicke theory reformulated in the Einstein conformal frame
\cite{FGN}, due to an explicit coupling between matter and the Einstein
frame scalar.} \cite{Wald}. The modified conservation equation (\ref{18})
is more transparent in the weak field limit, in which it reduces to 
\be\label{19} 
\frac{\partial\rho^{(m)}}{\partial t}+ \vec{\nabla} \cdot\left(
\rho^{(m)} \vec{v} \right) +\frac{2\kappa \xi \phi}{1-\kappa \xi \phi^2}
\left( \frac{\partial \phi}{\partial t} +\vec{\nabla} \phi \cdot \vec{v}
\right) \, \rho^{(m)} = 0 \; , 
\ee 
with $\vec{v}$ denoting the
three-dimensional velocity of the non-relativistic fluid; even in the
slow-motion limit, the usual energy conservation law $\partial\rho^{(m)}/
\partial t + \vec{\nabla} \cdot\left( \rho^{(m)} \vec{v} \right) = 0 $
acquires a correction. In the effective coupling approach, the total
energy density seen by an observer with four-velocity $u^c$ is 
\be \label{20} 
\rho_{(total)}^{(II)} \equiv T_{ab}^{(II)} u^au^b +
T_{ab}^{(m)} u^au^b = \rho_{ \phi}^{(II)}+\rho^{(m)} \;, 
\ee 
where 
\be \label{21} 
\rho_{\phi}^{(II)} =\left( u^c \nabla_c \phi \right)^2
+\frac{1}{2} \nabla^c \phi \nabla_c \phi + V( \phi) -\xi \Box ( \phi )^2
-\xi u^a u^b \nabla_{a}\nabla_{b} ( \phi^2 ) = \rho_{\phi}^{(I)} - \xi
\phi^2 G_{ab}u^au^b 
\ee 
or, using eq. (\ref{11}), 
\be \label{22}
\rho_{\phi}^{(II)}= \rho_{\phi}^{(I)}\left( 1-\kappa\xi\phi^2 \right)
=\kappa\xi\phi^2 \rho^{(m)} \; . 
\ee 
In the presence of ordinary matter,
$\rho_{\phi}^{(II)}$ is a mixture of $\rho_{\phi}^{(I)}$ and of
$\rho^{(m)}$ weighted by the factor $\kappa \xi \phi^2$.

\subsection{The mixed approach}

The third and last possibility to rewrite the field equations (\ref{2})
 is to bring the $\xi\phi^2G_{ab}$ term to the left hand side  but to 
keep the usual, constant, gravitational coupling, thereby obtaining
\be  \label{23}
G_{ab}=\kappa \left( T_{ab}^{(III)}+\frac{ T_{ab}^{(m)}}{1-\kappa\xi\phi^2}  
\right) \; , 
\ee
where 
\be \label{24}
T_{ab}^{(III)}=\frac{1}{1-\kappa\xi\phi^2} \,  \left[ \nabla_a \phi \nabla_b \phi -\frac{1}{2} g_{ab}
\nabla^c \phi \nabla_c \phi -V g_{ab} 
+\xi \left( g_{ab} \Box  -\nabla_{a}\nabla_{b} \right) ( \phi^2 ) \right] = 
\frac{T_{ab}^{(II)}}{1 -\kappa \xi \phi^2} \; ,
\ee
Clearly, the limitations due to division by the factor $\left( 
1-\kappa\xi\phi^2 \right)$ are present also in this mixed
approach. Note that
\be \label{25}
\kappa T_{ab}^{(III)}= \kappa_{eff} T_{ab}^{(II)}
\ee
and that the total energy density measured by an observer with
four-velocity $u^c$ is
\be \label{26}
\rho_{(total)}^{(III)} \equiv  T_{ab}^{(II)} u^au^b + 
T_{ab}^{(m)} u^au^b = 
\rho_{ \phi}^{(II)}+\frac{ \rho^{(m)}}{1-\kappa\xi\phi^2} \;.
\ee
In the absence of ordinary matter, $ \rho_{\phi}^{(III)}=\rho_{\phi}^{(II)}$.
This time the contracted Bianchi identities $\nabla^b G_{ab}=0$ yield
conservation of the {\em total} energy-momentum
tensor
\be   \label{27}
T_{ab}^{(total)}=T_{ab}^{(II)}+\frac{T_{ab}^{(m)}}{1 -\kappa \xi \phi^2} \;,
\ee
\be \label{28}
\nabla^b  T_{ab}^{(total)}=0 \;,
\ee
but $T_{ab}^{(III)}$ is not covariantly conserved:
\be   \label{29}
\nabla^b T_{ab}^{(III)}=
\frac{-2\kappa \xi \phi}{ \left( 1-\kappa \xi \phi^2 \right)^2} \, \left( \nabla^b \phi \right)
T_{ab}^{(m)}  \; .
\ee
However, in the absence of ordinary matter, $T_{ab}^{(III)}$ is conserved, 
$ \nabla^b T_{ab}^{(III)}=0$.

\subsection{Discussion}

The three possible cases are summarized in Table 1, and they trivially 
coincide for $\xi=0$. For $\xi<0$, the
scalar field $\phi$ does not possess the critical values (\ref{13}) and
the three approaches discussed above are mathematically equivalent, since
they differ only by legitimate algebraic manipulations of the field
equations (\ref{2}). However, they are physically inequivalent; in fact
the identification of the energy density, pressure, and effective equation
of state differs in the three approaches, and this is important in the
light of the present debate on energy conditions and the second law of 
thermodynamics with nonminimally coupled scalar fields as the source of 
gravity \cite{Hiscock,BarceloVisserCQG,VisserBarcelo,FordRomanpreprint}.
Moreover, the different conservation laws (\ref{8}), (\ref{15}) and
(\ref{29}) lead to different physical interpretations. 
For example, one encounters models of quintessence with NMC 
\cite{NMCquintessence} and observational constraints are often stated in 
terms of the effective equation of state. However, when the definitions 
of energy density and pressure are ambiguous, also the concept of 
equation of state becomes fuzzy.

In order to compare the different scalar field energy densities we assume,
in the rest of this paper, that $\phi$ is the only form of
matter. Besides the obvious simplification in eqs.~(\ref{22}) and
(\ref{26}), this situation is appropriate for the description of 
inflationary scenarios of
the early universe \cite{KolbTurner}, for late quintessence-dominated
cosmological scenarios \cite{NMCquintessence}, and for wormhole models
\cite{wormhole,Visserbook,BarceloVisserCQG}. 

For $\xi<0$, the effective coupling $\kappa_{eff} $ is always positive,
and $\rho^{(II)}_{\phi}$ has the same sign as
$\rho^{(III)}_{\phi}=\rho^{(I)}_{\phi}$. 

For $\xi>0$, in addition to the physical inequivalence, the three
approaches are also mathematically inequivalent, as discussed above, and
the scalar field possesses the critical values (\ref{13}).  If 
$|\phi|<
1/\sqrt{\kappa\xi}$, then $\kappa_{eff}>0$ and $\rho^{(II)}_{\phi}$ has
the same sign of $\rho^{(I)}_{\phi}=\rho^{(III)}_{\phi}$. If instead
$|\phi| > 1/\sqrt{\kappa\xi}$, the effective coupling $\kappa_{eff}$ is
negative, a regime that has been called ``antigravity'' and studied in the
literature \cite{Starobinsky}. In this regime sign$(
\rho^{(II)}_{\phi})=-$sign$( \rho^{(I)}_{\phi}) 
=-\mbox{sign} (\rho^{(III)}_{\phi})$.
However, $\kappa_{eff} \, \rho^{(II)}_{\phi}= \kappa \, \rho^{(I)}_{\phi}=
\kappa \, \rho^{(III)}_{\phi}$; while it is this product that
ultimately enters the field equations, the worries about the negative
energy fluxes associated to NMC and leading to time machines and
violations of the second law of thermodynamics \cite{FordRomanpreprint} are
expressed using $T_{ab}^{(III)}$ of our notations. Had one proceeded using
$ T_{ab}^{(II)}$ instead, one would have found, for $\xi>0$ and $|\phi|>
|\phi_c|$, {\em positive} energy density, but the same wormhole solutions.

The minimal conclusion that one draws from this discussion is that the
definition of energy density and flux must be carefully specified when
discussing energy conditions in the presence of NMC; and the situation in
this respect is now clarified.  Another relevant issue is whether there is
a physically preferred stress-energy tensor in the NMC theory described by
(\ref{1}). The answer is, to a certain extent, a matter of taste, and
depends on the task pursued; from a general point of view, the approach
\`a la Callan-Coleman-Jackiw presents certain advantages. First, the
corresponding stress-energy tensor $T_{ab}^{(I)}$ is always covariantly
conserved, even in the presence of ordinary matter $T_{ab}^{(m)}$; second,
this approach does not lead to loss of generality of the solutions. 
Third, there are
situations in cosmology (described in the next section) in which the
energy density $ \rho^{(I)}_{\phi}$ is automatically positive-definite,
while $ \rho^{(II)}_{\phi}$ is not. However, we cannot provide definitive 
arguments to rule out the effective and the mixed coupling approaches; we 
believe that such an argument cannot be presented, and the final 
decision on what approach is most convenient is left to the reader. 


\section{Energy conditions in FLRW cosmology}
\setcounter{equation}{0}

That the energy density of a nonminimally coupled scalar field can be
negative has been known for a long time (e.g.
\cite{Bekenstein,Madsen,Hiscock}), although the troublesome
consequences have only recently been emphasized
\cite{BarceloVisserCQG,FordRomanpreprint}. There are however situations 
in which negative energies do not occur with the definition of energy density
$\rho_{\phi}^{(I)}$ proposed as the physical one in the previous section.

Let us consider a Friedmann-Lemaitre-Robertson-Walker cosmology described
by the line element 
\be 
ds^2=-dt^2+a^2(t) \left[ \frac{dr^2}{1-Kr^2}+r^2
\left( d\theta^2 +\sin^2 \theta \, d\varphi^2 \right) \right] 
\ee 
in comoving
coordinates $\left( t,r,\theta,\varphi \right)$; it is assumed that the
dynamics of the universe are driven by the nonminimally coupled scalar
field (as is the case, for example, during inflation or in a late
quintessence-dominated era). The scale factor $a(t)$ and the scalar field
$\phi(t)$ satisfy the Einstein-Friedmann equations 
\be \label{31}
\frac{\ddot{a}}{a}=\dot{H}+H^2=-\, \frac{\kappa}{6} \left( 
\rho^{(I)}+3P^{(I)} \right)  
\ee 
\be \label{32} 
H^2=\frac{\kappa}{3}\, 
\rho^{(I)}-\frac{K}{3 a^2} 
\ee 
(where $P^{(I)}$ is the isotropic pressure
obtained from $T_{ab}^{(I)}$ for a time dependent scalar $\phi(t)$). The
Hamiltonian constraint (\ref{32}) implies that the energy density
$\rho^{(I)}_{\phi}$ is automatically non-negative for any solutions
$\left( a(t), \phi(t) \right) $ of the Einstein equations for spatially
flat ($K=0$) and for closed ($K=+1$) universes. This conclusion holds in
spite of the complicacy of the expressions for $\rho^{(I)}(t)$ and
$P^{(I)} (t)$, 
\be \label{33} 
\rho^{(I)}(t)=\frac{1}{2} \dot{\phi}^2 +3\xi
H^2\phi^2 +6\xi H\phi \dot{\phi} +V( \phi) \;, 
\ee 
\be \label{34}
P^{(I)}(t)= \left( \frac{1}{2}- 2\xi \right) \dot{\phi}^2 +2\xi H \phi
\dot{\phi} +2\xi ( 6\xi -1) \dot{H}\phi^2 +3\xi ( 8\xi -1) H^2 \phi^2
+2\xi \phi \frac{dV}{d\phi} \;, 
\ee 
from which nothing could {\em a priori}
be concluded.


\section{Negative energies} \setcounter{equation}{0}

The problem of negative energy fluxes with nonminimally coupled scalars
already appears with a flat background spacetime, as demonstrated by the
simple example of Ref.~\cite{FordRomanpreprint}. The negative energy flux 
problem
is by no means restricted to NMC \cite{FGN}; here we present an example,
analogous to that of Ref.~\cite{FordRomanpreprint}, in which weak 
gravitational
waves on a flat background in Brans-Dicke theory exhibit negative energies
on a macroscopic time scale. Scalar gravitational waves are currently
under investigation by many authors \cite{scalargw}.\footnote{
For a recent review of the status of gravitational wave detectors, see
\cite{GWD}; see also \cite {new2}.}

In the usual formulation of Brans-Dicke theory in the so-called Jordan
conformal frame the action is 
\be \label{35} 
S_{BD}=\int d^4x \, \sqrt{-g}
\, \left[ \phi_{BD} R -\frac{\omega}{2} \, g^{ab} \, \partial_a \phi_{BD}
\, \partial_b \phi_{BD} \right] +S_{(m)} \; , 
\ee 
which exhibits an explicit
coupling betweeen the Brans-Dicke field $\phi_{BD}$ and the Ricci
curvature.

Let us consider, in the weak field limit, scalar-tensor gravitational
waves around a flat background, described by 
\be \label{36}
g_{ab}=\eta_{ab}+h_{ab} \; , \;\;\;\;\;\;\;\;\;\;\;\;\;\;\;
\phi_{BD}=\phi_0 + \varphi \;, 
\ee 
where O$(h_{ab})=$O$( \varphi/\phi_0 )=$O$( \epsilon )$, where
$\epsilon$ is a small parameter.  
By linearizing the Brans-Dicke field equations in a region outside
the sources, one obtains (e.g. \cite{Will}) 
\be \label{37}
R_{ab}=\frac{\partial_{a}\partial_{b} \varphi}{\phi_0}\,
+\mbox{O}( \epsilon^2)  \equiv T_{ab}^{(J)} \left[ \varphi \right]+
\mbox{O}( \epsilon^2) \;, 
\ee 
\be \label{38} 
\Box \varphi=0 \;. 
\ee 
The
energy density of gravitational waves is given, to order $\epsilon$, only
by the scalar waves. The tensor modes give a contribution described by the
Isaacson effective stress-energy tensor \cite{MTW}, which is only of second 
order.
The energy density measurable by an observer with four-velocity $u^a$ is
\be \label{39} 
\rho^{(J)}=T_{ab}^{(J)} \left[ \varphi \right] u^a u^b
+\mbox{O}( \epsilon^2)  
\ee 
and its sign is indefinite. By taking, for
example, a monochromatic scalar wave $\varphi=\varphi_0 \cos \left( k_c
x^c \right) $, one has the oscillating density $\rho^{(J)}=-\left( k_c x^c 
\right)^2 \, \varphi / \phi_0 $. Hence, the
energy of gravitational waves emitted by a monochromatic source like a
binary system is negative over time intervals equal to half of the orbital 
period, which typically is  on the scale of days or months. It is hard to 
accept that a binary stellar system gains energy by {\em emitting}  
gravitational waves.

The root of the problem lies in the non-canonical form of the scalar field
energy density in Brans-Dicke theory (in its Jordan frame formulation used
so far): $T_{ab}$ has a part that is proportional to the second covariant
derivative $\nabla_a \nabla_b \phi_{BD}$, instead of being quadratic in
the first derivatives, $\nabla_a \phi_{BD} \, \nabla_b \phi_{BD}$.  
$\nabla_a \nabla_b \phi_{BD}$ is the only first order term that survives
in the weak field limit, giving the energy density (\ref{39}).  A
comparison with eq.~(\ref{5}) shows that this kind of terms also appears  
in the energy-momentum tensors $T_{ab}^{(I)}$, $T_{ab}^{(II)}$ and
$T_{ab}^{(III)}$ for a nonminimally coupled scalar. It is not surprising
that there is a common root for the energy problem in NMC and in 
Brans-Dicke theory, since the action (\ref{1}) can be rewritten as that
for a scalar-tensor theory (\ref{35}), with a $\phi$-dependent Brans-Dicke
parameter $\omega( \phi)$.

The usual way to cure the problem of negative energy fluxes is to
reformulate the theory in the Einstein conformal frame (see
Ref.~\cite{FGN} for a review); we illustrate the procedure for the
previous example of
 weak field Brans-Dicke gravitational waves.

One performs the conformal transformation of the metric
\be \label{40} 
g_{ab} \longrightarrow \tilde{g}_{ab}=\Omega^2 g_{ab} \;, \;\;\;\;\;\;\;\;\;
\Omega= \sqrt{G\phi} 
\ee 
and redefines the Brans-Dicke scalar according to
\be \label{41} 
\phi_{BD}\longrightarrow \tilde{\phi} =\left( \frac{2\omega
+3}{16\pi G} \right)^{1/2} \ln \left( \frac{\phi_{BD}}{\phi_0} \right) \;.
\ee 
The Einstein conformal frame is the set of variables $\left(
\tilde{g}_{ab}, \tilde{\phi} \right)$, as opposed to the Jordan conformal
frame $\left( g_{ab}, \phi_{BD} \right)$. In the Einstein frame,
scalar-tensor gravitational waves are described by 
\be \label{42}
\tilde{g}_{ab}=\eta_{ab}+\tilde{h}_{ab} \;, \;\;\;\;\;\;\;\;\;\;\;\;\;
\tilde{h}_{ab}=h_{ab}+\frac{\varphi}{\phi_0}\, \eta_{ab}\;,
\;\;\;\;\;\;\;\; \tilde{\phi}=\tilde{\phi}_0+\tilde{\varphi} 
\ee 
where,
according to eq.~(\ref{41}), 
\be \tilde{\varphi}=\left( \frac{2\omega +3}{16\pi G} \right)^{1/2}
\,\,\varphi \;. 
\ee In regions of spacetime
outside sources, the Einstein frame linearized field equations are
\be 
\tilde{R}_{ab}-\frac{1}{2} \tilde{g}_{ab} \tilde{R} =\kappa \left\{
\tilde{T}_{ab} \left[ \tilde{\varphi}\right] +T_{ab}^{(eff)} \left[
\tilde{h}_{cd} \right] \right\} \;, 
\ee 
\be 
\Box \tilde{\varphi}=0 \;, 
\ee
where \be \tilde{T}_{ab}\left[ \tilde{\varphi} \right] = \partial_a
\tilde{\varphi} \, \partial_b \tilde{\varphi} -\frac{1}{2}\, \eta_{ab}
\, \partial^{c}\tilde{\varphi} \, \partial_c \tilde{\varphi} 
\ee
and
$T_{ab}^{(eff)} \left[ \tilde{h}_{cd} \right] $ is Isaacsons's effective
stress-energy tensor \cite{MTW} associated to the tensor modes 
$\tilde{h}_{cd}$. In
the Einstein frame description, the energy momentum tensor of the scalar
field $\tilde{T}_{ab}\left[ \tilde{\varphi} \right] $ has the canonical
quadratic dependence on $\partial_d \tilde{\varphi}$, and both scalar and
tensor modes give contributions of the same order (that is, of second
order) to the field equations. For a monochromatic plane wave
$\tilde{\varphi} =\tilde{\varphi}_0 \cos \left( l^c x_c \right)$, one has
now \be \tilde{\rho} =\tilde{T}_{ab}u^a u^b =\left( l_a u^a
\tilde{\varphi}\right)^2 +T_{ab}^{(eff)} u^a u^b \geq 0 \;, \ee with a
positive-definite contribution from the scalar modes.


\section{NMC and gravitational waves}
\setcounter{equation}{0}

Due to the explicit coupling between the gravitational and the scalar
fields in the NMC theory (\ref{1}), when gravitational waves are excited,
also scalar waves are generated. We set $V( \phi )=0 $ and
$T_{ab}^{(m)}=0$.  The stress-energy tensor (\ref{5}) of $\phi$ contains
the term $ \xi \left( g_{ab} \Box - \nabla_a \nabla_b \right) ( \phi^2) $
linear in the second derivatives of $\phi$ instead of a canonical term
quadratic in its first derivatives $\nabla_c \phi$: this non-canonical
structure is analogous to that of the effective stress-energy tensor of
the Brans-Dicke scalar. Therefore, in principle, one could encounter here 
the same problems discussed for scalar-tensor gravitational waves in
Brans-Dicke theory. However, this is not the case, at least in the first
order perturbation analysis considered in this paper, as is shown below.

One expands the metric and the scalar field around their flat space
values, \be g_{ab}=\eta_{ab}+h_{ab} \;, \;\;\;\;\;\;\;\;\;\;\;
\phi=\phi_0+\psi \;, \ee with $\phi_0=$const. (solutions with constant
scalar field are discussed in the Appendix), and O$( h_{ab})=$O$( \psi) 
=$O$( \epsilon )$. It is easily seen from the
field equations (\ref{2}) that the requirement that spacetime be flat
implies $\phi=$constant. The offending term in $T_{ab}^{(I)}$ then reduces
to $\xi \left( g_{ab}\Box -\nabla_a \nabla_b \right) (\phi^2)=-2\xi \phi_0
\, \partial_a \partial_b \psi +\mbox{O}( \epsilon^2)$, where we used the
equation \be \Box \psi=0 +\mbox{O}( \epsilon^2) \;, \ee which follows from
the trace of the field equations 
\be 
R=-6\xi \phi_0 \kappa \, \Box \psi
+\mbox{O}( \epsilon^2) \;, 
\ee 
and from the Klein-Gordon equation 
\be 
\Box \phi -\xi R \phi=0 \;. 
\ee 
Hence, the negative energy problems seem  to be
present; however, in NMC theory, contrarily to Brans-Dicke theory, the
only value of the constant $\phi$ compatible with a flat background is
$\phi_0=0$. In fact, a constant non-vanishing scalar field corresponds to
Schwarzschild and anti-Schwarzschild solutions \cite{BarceloVisserCQG}. By
setting $\phi_0=0$ as required for a flat background, the troubles
disappear since $T_{ab}^{(I)} $ reduces, to first order, to a canonical
form quadratic in the first derivatives of the field, and the energy
density of scalar modes in NMC theory is positive definite.


\section{Discussion and conclusions} 
\setcounter{equation}{0}

We discussed the possible mathematical definitions of energy density and 
pressure of a nonminimally coupled scalar field, and the ambiguities 
arising when one studies the weak and null energy conditions for such a 
field. Three inequivalent definitions of $\rho$ and $P$ are possible.

>From a  fundamental point of view, the difficulty in identifying the 
``physically correct'' stress-energy tensor hints at the problem of the 
localization of gravitational energy. In fact, one can regard the 
physical system described by the action (\ref{1}) as two coupled 
subsystems, the gravitational tensor field $g_{ab}$ and the 
(non-gravitational) scalar field $\phi$, with the term $-\sqrt{-g} 
\xi\phi^2 R/2$ in the Lagrangian density as an explicit interaction 
term.

While no general prescription is known for the energy density of the 
gravitational field \cite{Wald,MTW}, a totally satisfactory definition 
of stress-energy tensor is available for a minimally coupled ($\xi=0$) 
scalar field. One should recognize that, when $\xi=0$, $g_{ab}$ and 
$\phi$  exchange energy and momentum during their dynamical evolution; as 
a consequence, the stress-energy tensor of $\phi$ contains terms 
describing this interaction. The impossibility of localizing the 
gravitational field energy may therefore be the source of the problems 
with NMC. Then, the reported violations of the 
second law of thermodynamics by NMC \cite{FordRomanpreprint} are not 
surprising since the scalar $\phi$ under study is not an isolated 
system, but rather a subsystem coupled to a second one, and we know 
nothing about the energy 
density and entropy of the latter. The second law refers to 
the isolated global system, not to a subsystem of it.

A similar problem affects other theories of gravity in which a scalar 
field explicitly couples to gravity: string theories, 
supergravity, Brans-Dicke theory, scalar-tensor and higher derivative 
theories; negative kinetic energy densities appear. This is problematic 
when the kinetic terms dominate the dynamics, as in the example of 
Sec.~4. In scalar-tensor theories of gravity the 
scalar field itself is a gravitational field,  and its stress-energy 
tensor contains a part of the gravitational energy density.

The usual cure for the problem has been the reformulation of the 
theory  in the Einstein conformal frame 
\cite{BD,FordRomanpreprint,FGN}. 
This way out of the problem leaves many authors dissatisfied because it 
involves a radical change of the theory, thereby losing much of its 
original motivation. We feel that a completely satisfactory answer to the 
problem cannot come without a successful solution of the 
gravitational energy localization problem in general relativity. 

It is at present unclear whether the view of NMC as the physics of two 
coupled  subsystems  may be useful in the quest for an 
approximate localization of gravitational energy.

\clearpage


\section*{Appendix: solutions with constant scalar field}

\def\theequation{A.\arabic{equation}}\setcounter{equation}{0}
\setcounter{equation}{0}

Let us adopt approach~I and assume that $T_{ab}^{(m)}=0$; then the
field equations are 
\be \label{A1} 
G_{ab}=\kappa T_{ab}^{(I)} =\kappa
\left[ \partial_a \phi \, \partial_b \phi -\frac{1}{2} \, g_{ab} 
\, \partial^c \phi \,
\partial_c \phi -V( \phi) \, g_{ab} +\xi \left( g_{ab}\Box -\nabla_a \nabla_b
\right)(\phi^2) -\xi \phi^2 G_{ab} \right] \;, 
\ee 
\be \label{A2} \Box
\phi -\xi R \phi-\frac{dV}{d\phi}=0 \;. 
\ee 
By assuming that
$\phi=\mbox{const.} \equiv \phi_0$, one has 
\be \label{A3} 
G_{ab} =-\kappa V_0
g_{ab} +\kappa \xi \phi_0^2 G_{ab} \;,
\ee 
\be \label{A4} 
\xi R \phi_0 +V'_0
=0 \;. 
\ee 
In the non-trivial case $\phi_0 \neq 0$, one has \be \label{A5}
R=-\,\frac{V_0' }{ \xi \phi_0} =\mbox{const.} \; , 
\ee 
i.e. all these  solutions
have constant Ricci curvature. Eq.~(\ref{A3}) then implies that \be
\label{A6} R=4\kappa V_0 +\kappa \xi \phi_0^2 R \ee and, if $\phi_0^2\neq
1/\kappa\xi$, \be \label{A7} R=\frac{4\kappa V_0}{1-\kappa\xi\phi_0^2} \;.
\ee By comparing eqs.~(\ref{A5}) and (\ref{A7}), one obtains the
constraint between  $V, \phi_0$, and $\xi$ 
\be \label{A8} 
\phi_0 V_0
=\frac{-V_0'}{4\kappa\xi \left( 1-\kappa\xi\phi_0^2\right)} \;. 
\ee 
If
instead $ \phi^2=\phi_c^2 \equiv 1/ ( \kappa\xi)$ with $\xi>0$, then it
must be $V_0=0$, $R=0$, and $V_0'=0$;  the potential $V( \phi)$ must have
a zero and horizontal tangent in $\pm \phi_c$. These critical scalar field
values have been studied in scalar field cosmology for the potential \be
\label{A9} V( \phi)= \frac{\alpha}{2} \phi^2 \left( \frac{6}{\kappa}
-\phi^2 \right) \ee and for $\xi =1/6$; in this case all the solutions
with constant Ricci curvature are classified \cite{GunzigetalCQG}. For
Lorentzian wormholes without potential $V$, all the solutions
corresponding to the critical scalar field values are also classified
\cite{BarceloVisserCQG}. 


\clearpage

\begin{flushleft}

\noindent
\begin{table}
{\small {\small 
\begin{tabular}{|c|c|c|c|}
\hline  
& 
$\begin{array}{c} \mbox{(I)}\\ \mbox{ approach \'{a} la CCJ} \end{array} $ 
& 
$\begin{array}{c} \mbox{(II)} \\ \mbox{effective coupling approach} 
\end{array} $  
& $\begin{array}{c} \mbox{(III)} \\ \mbox{mixed  approach} \end{array} $\\ 
\hline $ 
\begin{array}{c} \mbox{field}\\ \mbox{equations} \end{array} $  &  
$ G_{ab}=\kappa \left(  T_{ab}^{(I)}  +T_{ab}^{(m)} \right) $  & 
$   G_{ab}=\kappa_{eff} \left(  T_{ab}^{(II)}  +T_{ab}^{(m)} \right)   $
& $  G_{ab}=\kappa \left(  T_{ab}^{(III)}  
+\frac{T_{ab}^{(m)}}{1-\kappa\xi\phi^2} \right)   $\\
\hline  $ T_{ab} $  & 
$ \begin{array}{c}
T_{ab}^{(I)}=\partial_a \phi \partial_b \phi 
-\frac{1}{2} g_{ab} \partial^c \phi \partial_c \phi \\
+\xi \left( g_{ab} \Box -\nabla_a \nabla_b \right) ( \phi^2) \\
-Vg_{ab} +\xi  \phi^2 G_{ab} \end{array} $
& $ T_{ab}^{(II)}=T_{ab}^{(I)} -\xi \phi^2 G_{ab} $  & 
$ T_{ab}^{(III)}=\frac{T_{ab}^{(II)}}{1-\kappa\xi\phi^2}   $\\
\hline $\begin{array}{c} \mbox{conservation}\\
\mbox{ law} \end{array}$  & 
$ \nabla^b T_{ab}^{(I)}=0 $  &
$ \begin{array}{c} \nabla^b 
T_{ab}^{(II)}=\frac{-2\kappa\xi\phi}{1-\kappa\xi\phi^2} \cdot \\ 
\left( T_{ab}^{(II)} +T_{ab}^{(m)} \right) \nabla^b \phi \end{array} $ &
$ \nabla^b T_{ab}^{(III)}=\frac{-2\kappa\xi\phi}{1-\kappa\xi\phi^2}  
T_{ab}^{(m)} \nabla^b \phi $ \\
\hline $ \begin{array}{c} \mbox{energy}\\ \mbox{density} \end{array}$  & 
$ \begin{array}{c} 
\rho_{\phi}^{(I)}=( u^c \partial_c \phi  )^2+\frac{1}{2} \partial^c \phi 
\partial_c \phi  \\  
-\xi u^a u^b \nabla_a \nabla_b ( \phi^2) \\
+V -\xi \Box ( \phi^2) +\xi \phi^2 G_{ab} u^a u^b 
\end{array} $ &
$ \rho_{\phi}^{(II)}=\rho_{\phi}^{(I)} \left( 1-\kappa\xi\phi^2 \right) $&
$ \rho_{\phi}^{(III)}=\rho_{\phi}^{(II)} $  \\ 
\hline
\end{tabular}
\caption{a comparison of the three possible approaches to NMC.} 
}}
\end{table} 
\end{flushleft}

\end{document}